\documentclass[aps,pre,reprint]{revtex4-2}
\usepackage{amssymb}
\usepackage{amsmath,bm}
\usepackage{graphicx,color}
\usepackage{bbm}
\usepackage[bottom]{footmisc}
\usepackage{multirow}
\usepackage{mathrsfs}

\usepackage{hyperref}
\hypersetup{
    colorlinks = true,
    linkcolor = blue,
    anchorcolor = blue,
    citecolor = magenta,
    filecolor = blue,
    urlcolor = blue
}

\begin{document}
\title{Length-scale Dependence of Stokes-Einstien Breakdown in Active Glass-forming Liquids}
\author{Anoop Mutneja $^1$, and Smarajit Karmakar $^{1}$}
\affiliation{
$^1$ Tata Institute of Fundamental Research, 36/P, Gopanpally Village, Serilingampally Mandal,Ranga Reddy District,
Hyderabad, Telangana 500107, India
	}

\begin{abstract}
Stokes-Einstein (SE) relation, which relates diffusion constant with the viscosity of a liquid at high temperatures in equilibrium, is violated in the supercooled temperature regime. Whether this relation is obeyed in nonequilibrium active liquids is a question of significant current interest to the statistical physics community trying to develop the theoretical framework of nonequilibrium statistical mechanics. Via extensive computer simulations of model active glass-forming liquids in three dimensions, we show that SE is obeyed at a high temperature similar to the equilibrium behaviour, and it gets violated in the supercooled temperature regimes. The degree of violation increases systematically with the increasing activity which quantifies the amount the system is driven out of equilibrium. First passage-time (FPT) distributions helped us to gain insights into this enhanced breakdown from the increased short-time peak, depicting hoppers. Subsequently, we study the wave vector dependence of SE relation and show that it gets restored at a wave vector that decreases with increasing activity, and the cross-over wave vector is found to be proportional to the inverse of the dynamical heterogeneity length scale in the system. Our work showed how SE violation in active supercooled liquids could be rationalized using the growth of dynamic length scale, which is found to grow enormously with increasing activity in these systems.
\end{abstract}
\maketitle
\section{Introduction}

The theory of Brownian motion captures the random erratic motion of a probe suspended in a host, and at its heart is the assumption of the separation of time scales. Put simply, the Brownian particle is hit by smaller particles in the medium (smaller time scale), which creates a non-zero force causing it to move (larger time scale). However, for the particle to move, the host particles must displace through viscous dynamics, facilitating the Brownian dynamics of a probe particle. Einstein \cite{Einstein1905} linked the probe and host dynamics by equating the two-time scales, namely the viscous time scale, $\eta R^3/T$, and the diffusive time scale of the probe $R^2/D_P$, via fluctuation-dissipation relation, up to a constant. It is known as the Stokes-Einstein (SE) relation \cite{HANSEN2013265} and is obeyed by a probe if $\eta D_p/k_BT$ is independent of temperature. Perrin \cite{Perrin} used the SE relation to determine Avogadro's number for the first time. This probe can be any external particle, or the particle of the medium itself, leading to the self version of the SE relation,  $\eta D_s/k_BT=constant$. 

Supercooled liquids, which show a considerable viscosity increase with a slight decrease in temperature, challenge the SE relation \cite{Liu2015, Adhikari2021,Rajian2007,Swallen2003,Parmar2017, Sengupta2014, Bhowmik2016,Ediger2012,Berthier2011}.
The decrease in diffusion constant on reducing temperature does not compensate for the massive increase in the system's viscosity, resulting in Stokes-Einstein breakdown (SEB). Both experiments and numerical studies have demonstrated that the SE parameter ($\eta D_s/k_BT$) increases as temperature decreases. The SE breakdown is also a hallmark of dynamic heterogeneity (DH), the existence of spatial regions with different mobilities. The broadening of relaxation time distribution would result in the bifurcation of the two-time scales; the slower set would control the viscous time scale, while the faster set would control the diffusive time scale; hence the violation \cite{Ediger2000}. Another set of reasoning involves the change of inherent particle dynamics; with increasing supercooling, the dynamics of the host particles change from normal viscous to collective, activated, and hence slower dynamics \cite{Faupel2003, Zhang2017, Mei2021}. SE relation works if the host follows the normal viscous dynamics; however, the collective, or the ``hopping dynamics" in the supercooled regime, would enhance the diffusion of the probe \cite{Faupel2003, Zhang2017, Mei2021}.

In \cite{Sengupta2014, Bhowmik2016}, the distribution of diffusion constants were explicitly shown to have a bimodal structure with increasing supercooling. The slower group of particles obeys the SE relation even in the supercooled regime if one compares their respective relaxation time and diffusion constant. In contrast, for the bulk measurement, relaxation time is controlled by the slower particles, and diffusion is controlled by the faster set, leading to the violation of SE. Alternatively, the system tends to follow a fractional SE relation, viz. $D_s \sim \tau_\alpha^{-(1-\omega)}$, with $\omega\in[0,1]$, note that the relaxation time $\tau_\alpha$ is usually used as a proxy for $\eta/T$. The deviation of $\omega$ from zero characterizes the degree of deviation and depends on the dynamical features of the system. In Ref. \cite{Bhowmik2016}, the SE relation was tested with increasing random pinning in the system; the power $\omega$ was shown to increase with increasing pinning concentration. In Ref.\cite{Adhikari2021}, SE relation is tested across the spatial dimensions ($d$) of the system, and it was noted that SE violation gets systematically reduced with increasing spatial dimensions and SE is completely recovered at the upper critical dimension $d_u \simeq 8$. Above this dimension, the system becomes more mean-field-like, and one expects diminished or negligible dynamic heterogeneity. In this study, our goal is to test the SE relation of a model active glass-forming liquid with increasing activity, which is introduced by giving a self-propulsion force to a fraction of particles in the system, as discussed in the model and methods section.
     
The intensity of SEB varies depending on the type of probe used \cite{Rajian2007,Liu2015}. For example, in a molecular liquid like othro-terphenyl (OTP) \cite{Liu2015}, the SE relation holds for larger probes (viz. rubrene), but SEB intensity increases as the probe size decreases. This suggests that there may be a specific length scale that controls SE breakdown. This idea has been tested through molecular dynamics simulations in Ref. \cite{MutnejaR, MutnejaT}, where probes of varying sizes were used in supercooled liquids. The growing, dynamic length scale, $\xi_d$ (introduced later), was found to control the SE breakdown. Additionally, it is well-established that the SE breakdown is not observed when probed at a smaller wavevector (large length scale) \cite{Parmar2017}. In this study, the authors checked the SE relation at different length scales by using the relaxation times obtained from the self-part of the intermediate scattering function ($F_s(k,t)$) (defined later) for different wavevectors $k$. A method was then proposed to obtain the SEB temperature for different wavevectors, which followed the dynamic length scale of the system.  

The growing length scale seems to be an essential requisite for the hypothesized glass transition, and it was found that at least two length scales are required to explain the hallmarks of glass-forming liquids \cite{Karmakar2014, KDSPNAS2009, KDSROPP, Tah2021}. Although there are alternative theories that considers glass transition to be purely dynamic in nature and does not necessarily require a growing static correlation length \cite{Schweizer2005,Reichman2005,Berthier2010}, but existence of a dynamical length scale is largely accepted across these diverse theories of glass transition.
This first scale is associated with DH patches, which control the SE breakdown, Non-Gaussian dynamics, and non-exponential relaxation. However, this scale does not causally account for the growing relaxation time of the system. The second length scale, known as the static length scale, was proposed in the RFOT framework \cite{RFOT1, RFOT2} to relate the slow dynamics causally. A few procedures are used to obtain these length scales numerically, such as the finite size scaling analysis \cite{Karmakar2014, KDSPNAS2009, KDSROPP} and the Block analysis \cite{Block}. 

The computer glass forming systems only fall short of the order of growing length scale; they only show a slowdown of 4-5 decades, with the length scale growing by a few particle diameters. Active supercooled systems have recently gained much attention because of the enormous growing length scales \cite{PaulD2021} and ubiquitous presence. Researchers in the past decade have focused on the emerging nonequilibrium class of soft matter systems, namely active glasses \cite{PaulD2021, PaulS2021,janssen2019,vijay2007,cugliandolo2019,caprini2020,merrigan2020, chaki2020, activerfot, szamel2016, saroj2018,  activemct, berthier2013} to develop an understanding of the nonequilibrium steady states and their statistical properties. The constituent particles, along with the glassy disordered behavior, are energy-driven (internal or supplied externally). The intercellular dynamics, wound healing, progressing cancer cells, and driven granular matter \cite{Malinverno2017,Cerbino2021,PhysRevLett.113.025701} are a few examples of active glassy systems. As already mentioned, such systems show enhanced DH, which is also experimentally sound; the cancer cells (more active) show greater DH than the healthy cells, which may lead to metastatic cell clustering \cite{Giuliano2018, Palamidessi2019}. The dynamic length scale in these systems can grow by as much as $30$ particle diameters (for the largest studied activity), providing a perfect playground to test various tools developed for equilibrium systems to work in such superlative nonequilibrium systems.

This work addresses the length-scale variation of SE violation in the active system. The reason behind its breakdown in a passive supercooled system is debated; one stream of thought links it with the DH, while the other thinks it to be an effect of change in inherent dynamics, from viscous to activated. A natural question that arises in active systems is whether SE relation is obeyed or not. As active systems are in nonequilibrium states, the SE breakdown need not require any reason. Nevertheless, the SE relation at high temperatures seems to be obeyed perfectly like any other liquid. Thus validity or violation of SE relation in active supercooled liquid is a pertinent question, and if it is violated, whether the degree of violation increases with increasing activity and whether dynamic heterogeneity plays any role in this context or not are questions worth further studies. These questions are non-trivial as we are asking these questions for a system that is at a nonequilibrium steady state, and it is apriori not known whether concepts of equilibrium statistical mechanics can be applied to these systems. In this work, we show that SE violation in active glass-forming liquids increases with increasing activity, and the physics of SE violation in active glasses can be understood in a unified manner using the growing dynamic heterogeneity length scale in the same way as in equilibrium passive glass-forming liquids. In fact, we show that a systematic wave vector dependence of SE relation can itself yield the dynamic heterogeneity length scale.


The rest of the article is organized as follows. We first give the details of the model active system we studied using extensive molecular dynamics simulations. Then we explain how various dynamical quantities are extracted using appropriate correlation functions. In the results section, we discuss the violation of the SE relation and how the SE relation gets restored once we look at it at a different probing length or wave vector with changing activity and temperatures in the system. We then show how the dynamic heterogeneity length scale intimately relates to the systems' SE violation. Finally, we conclude with implications of our findings on the physics of glass transition in nonequilibrium systems and their similarity and difference with the equilibrium counterparts. 

\begin{figure*}[htpb]
\centering
\includegraphics[width=1.0\textwidth]{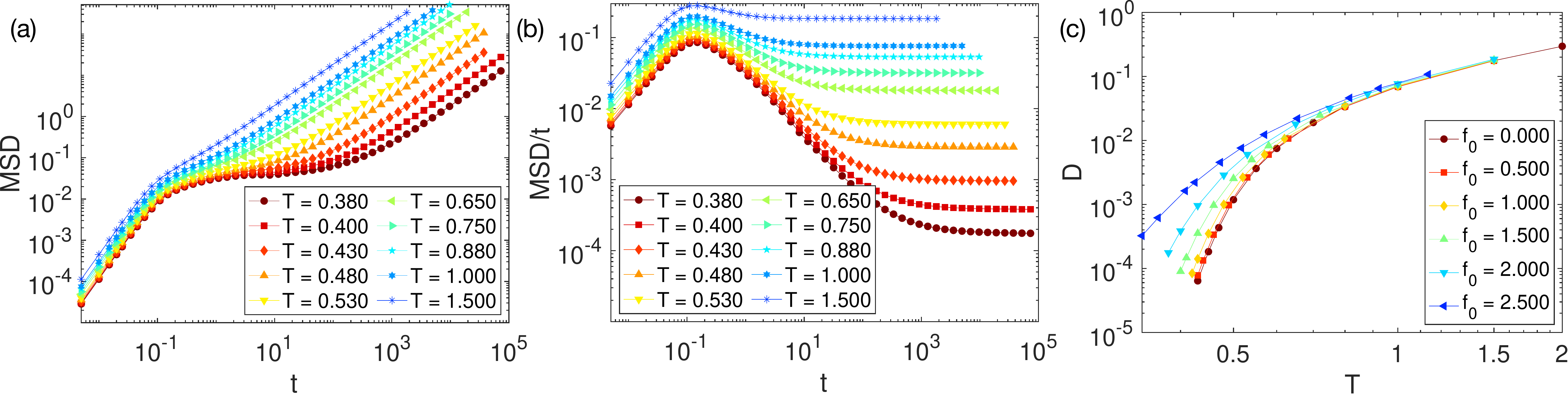}
\caption{(a) Time evolution plot of mean square displacement (MSD) for an active system with $f_0=2.0$. (b) The $MSD/t$ plot shows the eventual reaching of the diffusive behavior, with the diffusion constant proportional to the large time limit of the y-axis. (c) The extracted diffusion constant $D$ is plotted against the temperature for the systems with different activities.}
\label{MSDch6}
\end{figure*}
\section{Models and Methods}
This work consists of large-scale molecular dynamics simulations performed on the binary system (particle tag A and B) in the ratio of $80:20$, known in the literature as the Kob-Andersen model \cite{KA}, in three dimensions with the system size of $N=50000$ particles. The system is simulated in an NVT ensemble using a 3-chain Nose-Hoover thermostat. Activity is introduced as an eight-states clock model with the active force assigned as 
\begin{equation}
F^A_i=f_0 \left(k^i_x \hat{x} + k^i_y \hat{y} + k^i_z \hat{z} \right),
\label{activeforce}
\end{equation}
on $c=0.1$ concentration of randomly tagged active particles. In Eq.\ref{activeforce}, $ k^i_\alpha\in\pm1$ are chosen randomly after the persistence time $\tau_P=1.0$. $\sum_{i=1}^{N_A}k_\alpha^i=0$ is ensured for each spatial dimension to maintain the system's zero center-of-mass velocity. We averaged our results with $16$ statistically independent simulation runs for each state point, and the random particle choices are different for each run.

The obtained trajectory of the system is then analyzed to get the average mean squared displacement (MSD) (Fig.\ref{MSDch6}(a)) and hence the diffusion constant as $D\propto MSD/t$ (Fig.\ref{MSDch6}(b, c)) in the long time limit. The intermediate scattering function calculated as the Fourier transform of the density-density correlation function, calculated for different wavevectors $k_\alpha=2n_\alpha\pi/L$, gives the relaxation profile at different length scales. Here $\alpha$ is for dimensions, and $n_\alpha$ is a positive integer. $L$ is the system size. The self part of the intermediate scatter function, $F_s(k,t)$ is defined as
\begin{equation}
F_s(k,t)=\frac{1}{N}\left\langle\sum_{i=1}^Ne^{-i\textbf{k}.(\textbf{r}_i(t)-\textbf{r}_i(0)}\right\rangle,
\label{FsqtAct}
\end{equation}
where $\textbf{r}_i(t)$ is the instantaneous position of particle $i$ at time $t$ and the $\langle\cdots\rangle$ refers to thermal averaging. 
\begin{figure}[!htpb]
\centering
\includegraphics[width=0.5\textwidth]{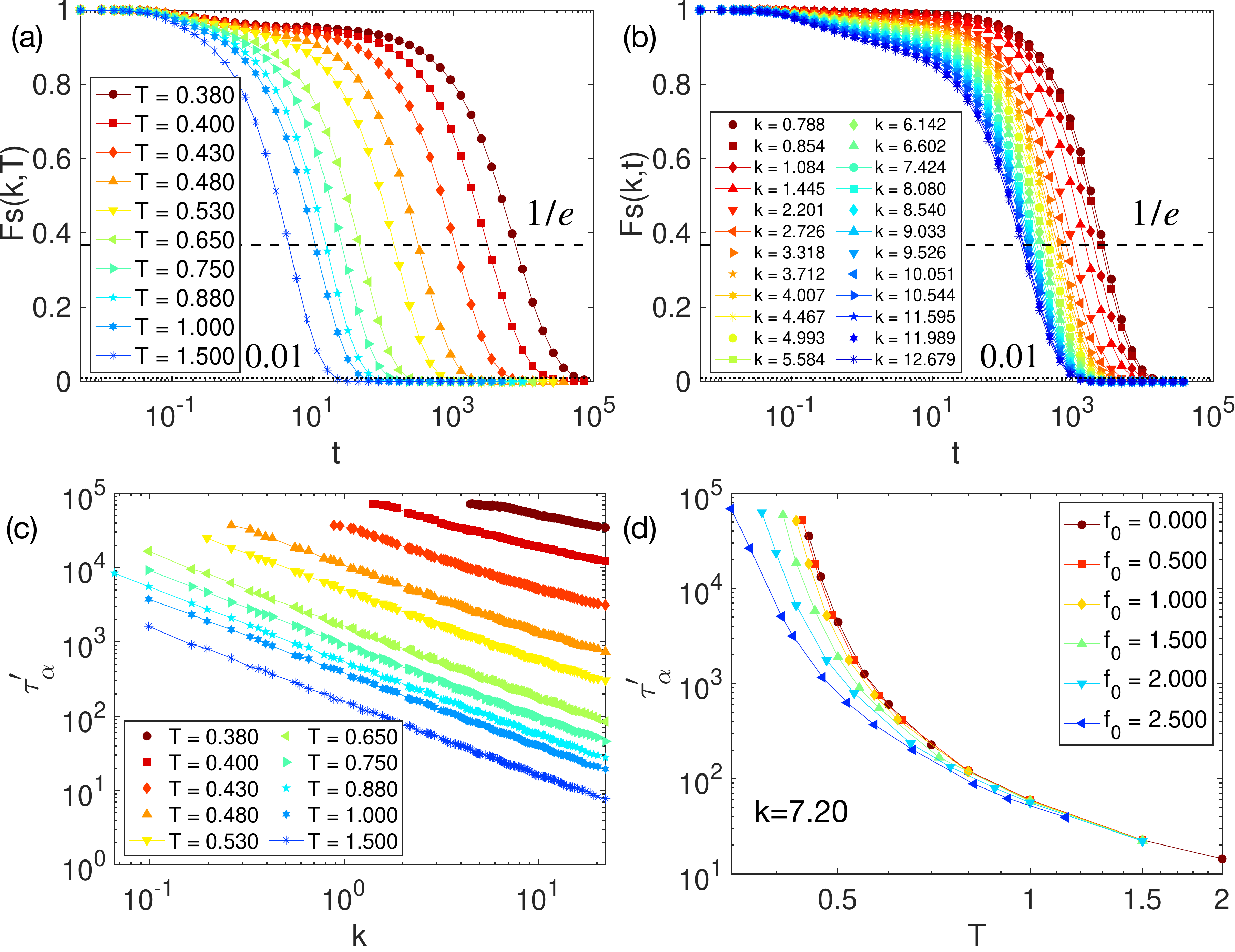}
\caption{(a) $F_s(k,t)$ plots for active system ($f_0=2.0$) subjected to different temperatures, showing the typical supercooled two-step non-exponential profile for low temperatures. These are the plots for $k=7.20$, where the structure factor peaks. (b) Wavevector variation of the $F_s(k,t)$ is shown for the same active system at temperature $T=0.480$. The relaxation profile becomes more and more exponential with decreasing $k$. (c) The extracted $\tau_\alpha^\prime(k,T)$ from $Fs(k,\tau_\alpha^\prime)=0.01$ cut is plotted against $k$ for different wavevectors. (d) The system's temperature variation of $\tau_\alpha^\prime$ ($k=7.20$) is shown for different activities. }
\label{Fsqtch6}
\end{figure}
Fig.\ref{Fsqtch6}(a) shows the $Fs(k,t)$ for an active system with $f_0=2.0$ for different temperatures, while the wavevector dependence can be seen in Fig.\ref{Fsqtch6}(b) for the same system with $f_0=2.0$, and $T=0.480$. The relaxation time $\tau_\alpha(k)$ for a wavevector $k$ is calculated as $F_s(k,\tau_\alpha(k))=1/e$. It has been shown in Ref.~\cite{Parmar2017} that with $\tau^\prime_\alpha$ calculated using $F_s(k,\tau^\prime_\alpha(k))=0.01$, does not show SEB for all $k$ in the high-temperature regime, in contrast of the $\tau_\alpha(k)$. Thus, we would use $\tau_\alpha^\prime$ in the rest of the article. The relaxation time of the system $\tau_\alpha^\prime(T)$ is calculated for wavevector $k=7.20$ at which the structure factor, $S(k)$, peaks. The static structure factor is defined as 
\begin{equation}
S(k)=\frac{1}{N}\left\langle\sum_{i,j=1}^N e^{-i\textbf{k}.(\textbf{r}_j-\textbf{r}_i)}\right\rangle.
\end{equation} 
The $k$ dependence of $\tau_\alpha^\prime$ can be seen in Fig.\ref{Fsqtch6}(c), while its temperature dependence for systems with different activities is shown in panel (d) of Fig.\ref{Fsqtch6}. Also, we only use the dynamic information of particles of type A.

\begin{figure*}[!htpb]
\centering
\includegraphics[width=1\textwidth]{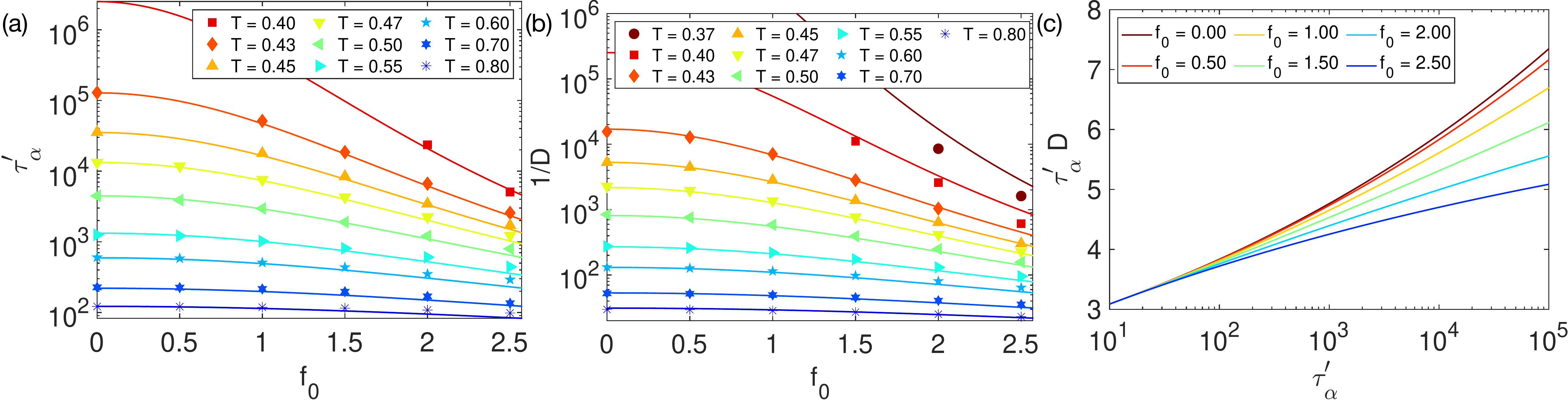}
\caption{(a) The relaxation times ($\tau_\alpha^\prime$) are shown for different temperatures as a function of active forcing $f_0$. The lines are the VFT lines and not fits. The equation used is $\tau_\alpha^\prime(T)=\tau_0exp\left(\frac{A}{T_{eff}-T_{VFT}}\right)$, where $T_{eff}=T+Kf_0^2$,  is the effective temperature with $K=0.015$ as a constant. The value of $T_{VFT}$ is obtained from the VFT fit of the passive case, along with the other unknown constants ($\tau_0$, $A$). Changing $T$ is what creates all the lines on the graph. (b) The graph also shows the inverse diffusion constant, further supporting the idea that active systems behave similarly to passive systems at higher effective temperatures. (c) We plot the SE parameter using VFT lines for different levels of activity in systems. The SE breakdown for $f_0=0$ is accurately detected, but it shows a weaker breakdown for higher activity levels. This is in contrast to what is numerically observed.} 
\label{VFT}
\end{figure*}
\section{Results}	
In Ref.~\cite{PaulD2021, activerfot,Mandal2016}, it was shown that relaxation behaviour as quantified by the relaxation time of an active system with changing activity could be completely understood using a suitably defined effective temperature, $T_{eff}$. Thus, it is natural to expect the diffusion constant to follow the same behaviour. In Fig.\ref{VFT}(a), we plot $\tau_\alpha^\prime$ as a function of $f_0$ for all 
the temperatures and the lines are the theoretical predictions within an effective temperature description as shown in Ref.\cite{activerfot}. In our case, we fit the relaxation time vs. temperature for the passive system ($f_0 = 0$) with Vogel-Fulcher-Tamman (VFT) formula as  
\begin{equation}
\tau_\alpha^\prime(T)=\tau_0exp\left(\frac{A}{T-T_{VFT}}\right)
\label{VFT_Eq}
\end{equation}
 and extract the unknown constants $\tau_0$, $A$, and $T_{VFT}$ and then used the following definition of effective temperature $T_{eff} = T + Kf_0^2$ and plotted the function  $\tau_\alpha^\prime(T)=\tau_0exp\left(\frac{A}{T_{eff}-T_{VFT}}\right)$ for all the value of $f_0$ studied in this work. The predictions (the solid lines) with $K = 0.015$ \cite{PaulD2021} are in perfect agreement with the simulation results (symbols). A similar analysis for inverse diffusion constant is done in Fig.\ref{VFT}(b), where we kept $T_{VFT}$ and $K$ to be the same and obtained the two free parameters $D_0$ and $B$ of the following equation
\begin{equation}
\frac{1}{D(T)}=\frac{1}{D_0}exp\left(\frac{B}{T-T_{VFT}}\right)
\label{VFT_EqD}
\end{equation}
using the diffusion constant data for the passive system only. The comparison with effective temperature description is reasonably good at higher temperatures but not so good at lower temperatures and higher activity,  indicating a possible decoupling between relaxation time, which is related to multiparticle correlation, and the diffusion constant, which is the average single-particle property of the system.

When we use the estimated diffusion constant and relaxation time within the effective temperature description (lines in Figs.\ref{VFT}(a) and (b)) and plot the SE parameter against relaxation time, we observe a violation of SE relation. However, the dependence on activity is not accurately represented in Fig.\ref{VFT}(c). It suggests a weaker violation in active systems with increasing activity, while stronger violations are observed in more active systems. Details about this will be discussed further below.

\begin{figure}[htpb]
\centering
\includegraphics[width=0.5\textwidth]{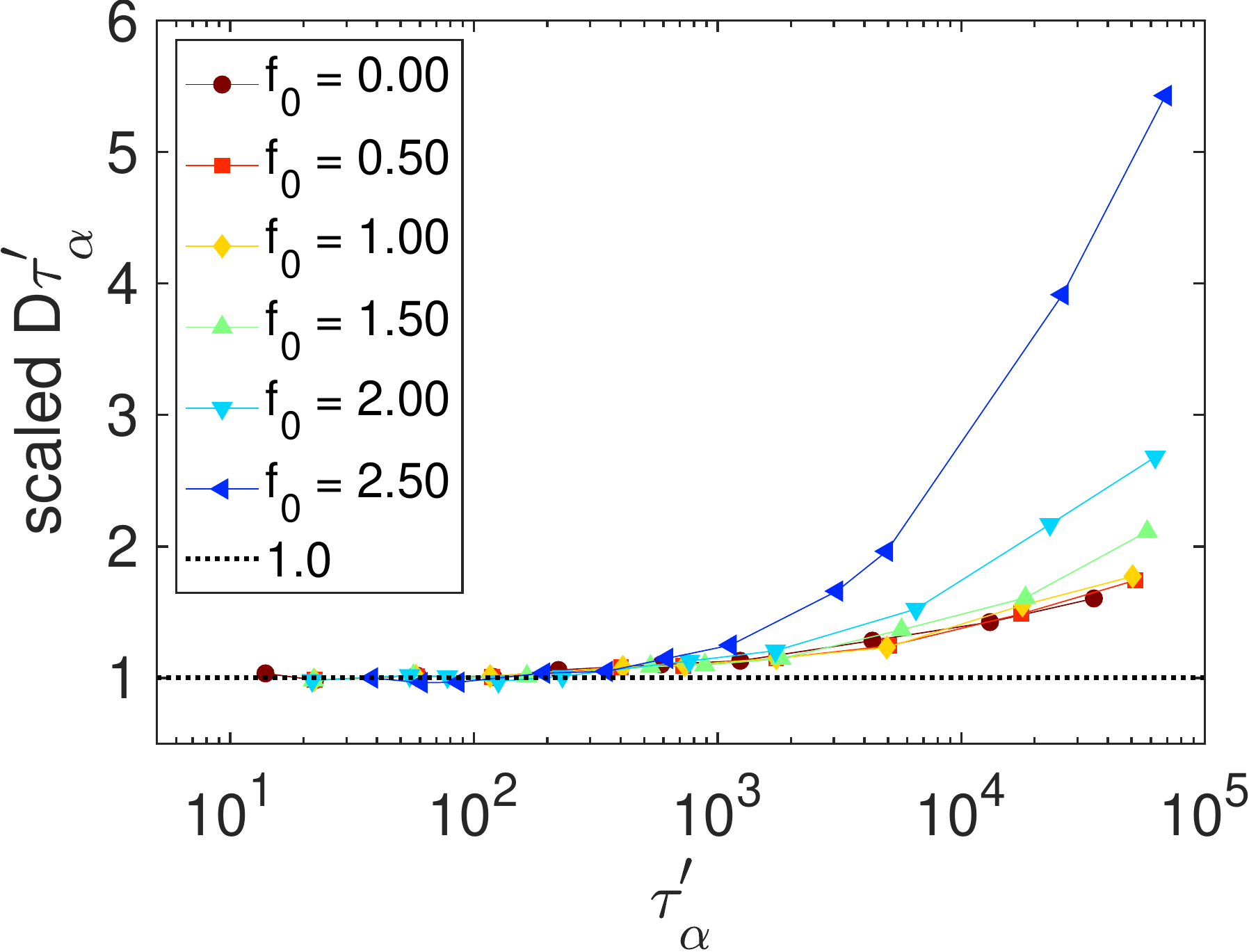}
\caption{The scaled SE parameter ($D\tau^\prime_\alpha$) is plotted against the relaxation time for systems with different activity strengths. The breakdown intensifies with increasing activity for a system with similar relaxation time, with the inference of increased DH. Also, the SEB happens at lower relaxation times for the systems of a larger activity. } 
\label{SEParamch6}
\end{figure}
\begin{figure*}[htb]
\includegraphics[width=\textwidth]{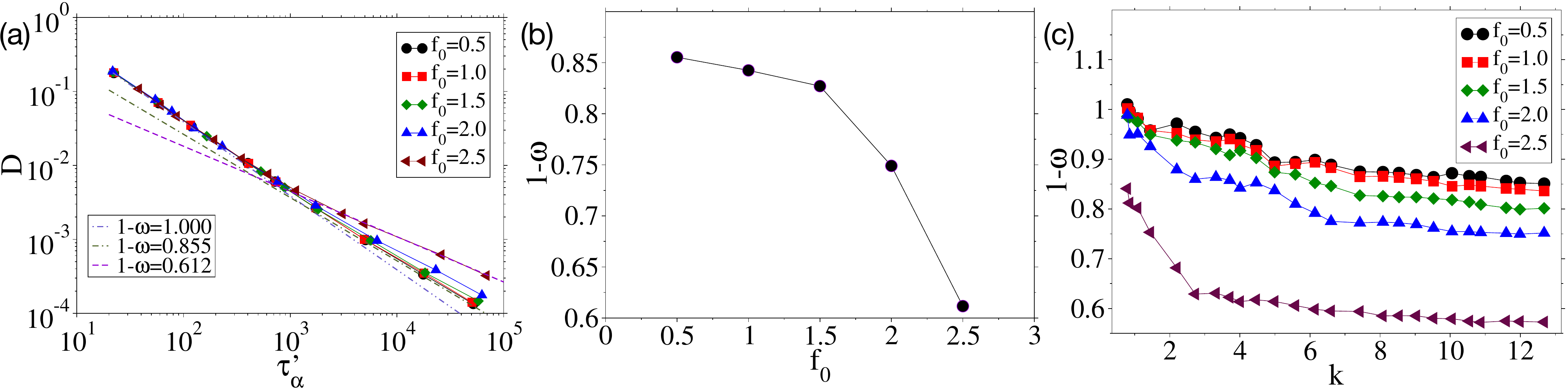}
\caption{(a) The cross plot of Log-log plot of the diffusion constant $D$ and $\tau^\prime_\alpha$, showing the SE relation for high temperatures, and fractional SE relation $D\propto \tau^{\prime - (1-\omega)}_\alpha$ in the low-temperature regime. With increasing activity, the power $1-\omega$ decreases further and further from one and can be seen in panel (b). (c) Wavevector variation of power $1-\omega$ for systems of different activities. $\omega$ approaches zero in the limit of small wavevectors and low temperatures, implying the soundness of SE relation.  }
\label{FractionalSE}
\end{figure*}
In Fig.\ref{SEParamch6}, we have plotted the SE parameter $D\tau^\prime_\alpha$ for systems with different active forcing $f_0$; all these systems obey the SE relation in the high-temperature limit and show the violation in the supercooling regime. The breakdown parameter escalates with decreasing temperature, while the rate of increase magnifies with increasing activity. The relaxation time where this breakdown happens also decreases with increasing activity. It is important to highlight that a similar effect was also seen with increasing random pinning concentration in the system \cite{Bhowmik2016}. It is clear from Fig.\ref{SEParamch6} that for systems with similar relaxation times but different activities, SE violation is much stronger for more active systems, which would have larger DH \cite{PaulD2021} leading to a strong link between DH and the SE breakdown. A substantial increase of DH with increasing activity has also been reported in  Keta et al. \cite{KetaPRL} and in some biological systems \cite{Giuliano2018, Palamidessi2019}.

Another way to characterize the observed intensified SE violation in an active system is to study the fractional SE relation, $D\tau^{\prime(1-\omega)}_\alpha$. The power $\omega=0$ corresponds to the SE relation, while for the supercooled liquids at low temperatures, $\omega>0$ is observed. The microscopic understanding of the fractional SE exponent is still unknown, but it depends on the microscopic interactions' details. In Fig.\ref{FractionalSE}(a), we have shown a log-log plot of the diffusion constant and relaxation time; at the high-temperature regime, all systems with different activities follow the SE relation ($\omega=0$), even with the same proportionality constant. However, the curves get separated from each other in the low-temperature limit; more active systems have a more significant deviation from $\omega=0$. Fig.\ref{FractionalSE}(b) shows the values of $\omega$ with increasing activity. To study the effect of probing wavevector, we also look at the SE violation with wavevector dependent relaxation time, $\tau_\alpha^\prime(k)$ using the following SE relation
\begin{equation}
D \sim \tau_\alpha^\prime(k)^{-(1-\omega(k))}. 
\end{equation}
Fig.\ref{FractionalSE}(c) shows the wavevector dependence of the power $\omega$ for systems of different activities in the limit of low temperatures. The power converges to $\omega=0$ in the limit of small $k$ for all activities; a systematic trend of decreasing $\omega$ with activity can also be seen. 

\begin{figure*}[htpb]
\includegraphics[width=1.\textwidth]{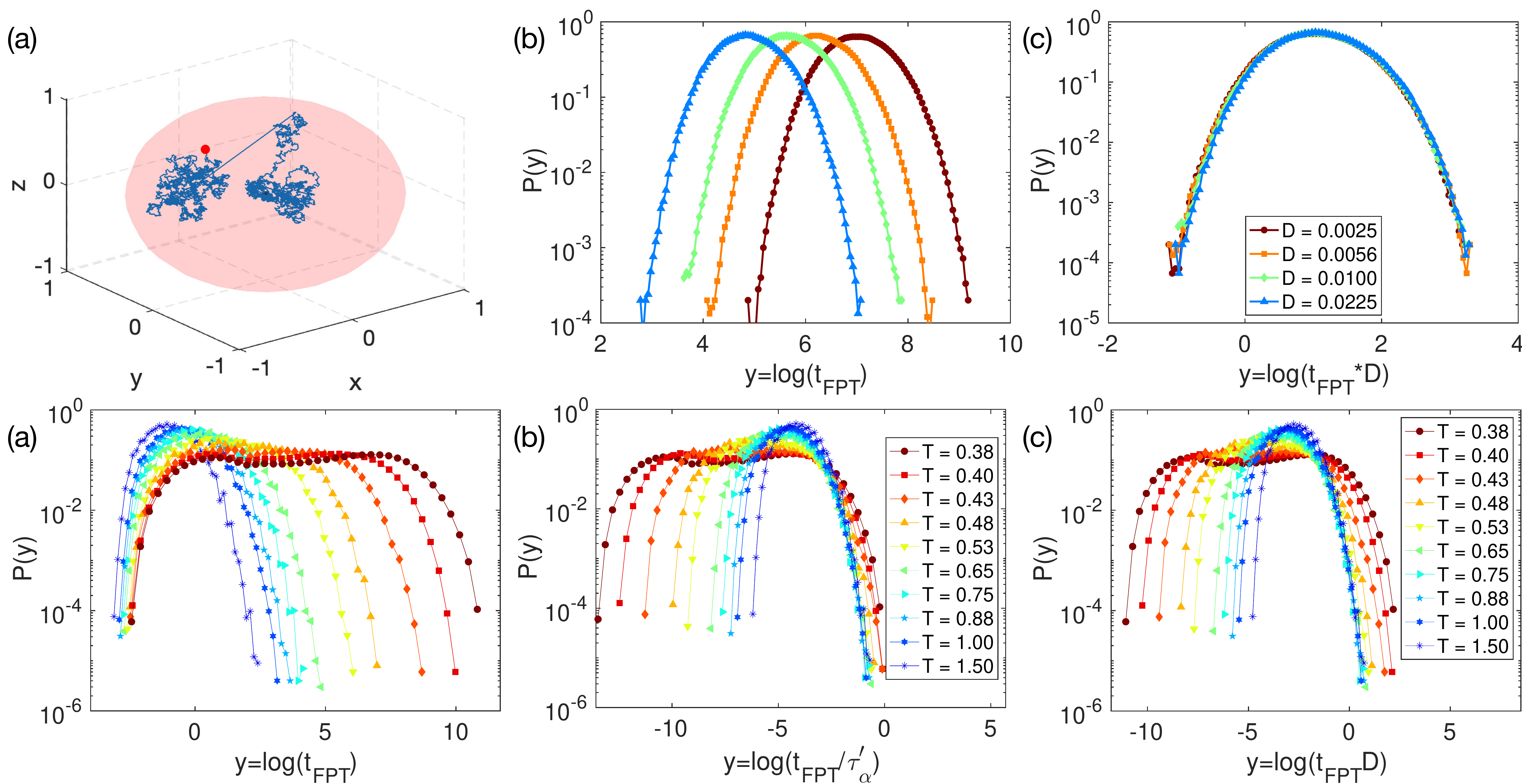}
\caption{(a) The Brownian particle starting from the origin in a space with an absorbing boundary at $|r|=1$; the time it gets absorbed would be the first passage time ($t_{FPT}$). (b) The distribution of FPT for such a Brownian particle with different diffusion constants $D$. (c) Simple scaling of $t_{FPT}*D$ brings all the curves to a master plot. (d) The FPT distributions for 3dKA active system with $f_0=2.0$ subjected to different temperatures. The appearance of a bimodal character and small time peak with increasing supercooling is proof of the DH and a reason for SE breakdown. Panel (e,f) shows the scaling of $t_{FPT}$, by $\tau_\alpha^\prime$, and $1/D$, respectively. The first one tends to give better scaling of large-time distribution than the latter, proving the involvement of different particles in calculating $\tau_\alpha^\prime$ and $D$.  }
\label{FPTF02.0}
\end{figure*}

To unearth the intricate role of activity on SE relation, we further look at the first-passage-time (FPT) of the particles. For a Brownian particle in 3d, the $FPT(r_c)$ is defined as the time when a particle crosses a sphere of radius $r_c$ for the first time starting from origin (see Fig.\ref{FPTF02.0}(a)). The Brownian particle in such an enclosed boundary would result in exponentially decaying FPT distributions. A typical FPT distribution would look like Fig.\ref{FPTF02.0}(b) for different choices of diffusion constant of the Brownian particle. All these curves can be scaled by the diffusion constant $D$ itself, as shown in Fig.\ref{FPTF02.0}(c), suggesting only one timescale in the Brownian problem with a homogeneous medium. Fig.\ref{FPTF02.0}(a-c) are the FPT plots for non-interacting Brownian particles, while Fig.\ref{FPTF02.0}(d-f) are the FPT plots of the supercooled system at different temperatures with active forcing of $f_0=2.0$. With increasing supercooling, the FPT distribution shows a bimodal behavior (Fig.\ref{FPTF02.0}(d)), with the emergent peak at short times being the contribution from the hoppers in the systems. This short-time peak systematically increases with increasing supercooling. If we try to scale the FPT with the relaxation time of the system, Fig.\ref{FPTF02.0}(e) (assuming the SE relation), the large-time part scales to some extent to a master plot; however, the hoppers fall out. On the other hand, the scaling with the diffusion constant, Fig.\ref{FPTF02.0}(f), does not show such collapse. The broadening of FPT distribution, especially in the short time regime, indicates the existence of a population of particles having diffusion constants that are significantly different from each other and is one of the primary causes for the SE breakdown. This suggests that distinct sets of particles with different diffusion constants contribute to estimating the relaxation time $\tau_\alpha^\prime$ and mean diffusion constant of the system, $D$. The slower set of particles appears to control $\tau_\alpha^\prime$ as the right-hand part of the FPT distribution can be scaled to a large extent by the average relaxation time of the system, while $D$ gets contributions from all the sets.

\begin{figure}[htpb]
\includegraphics[width=.5\textwidth]{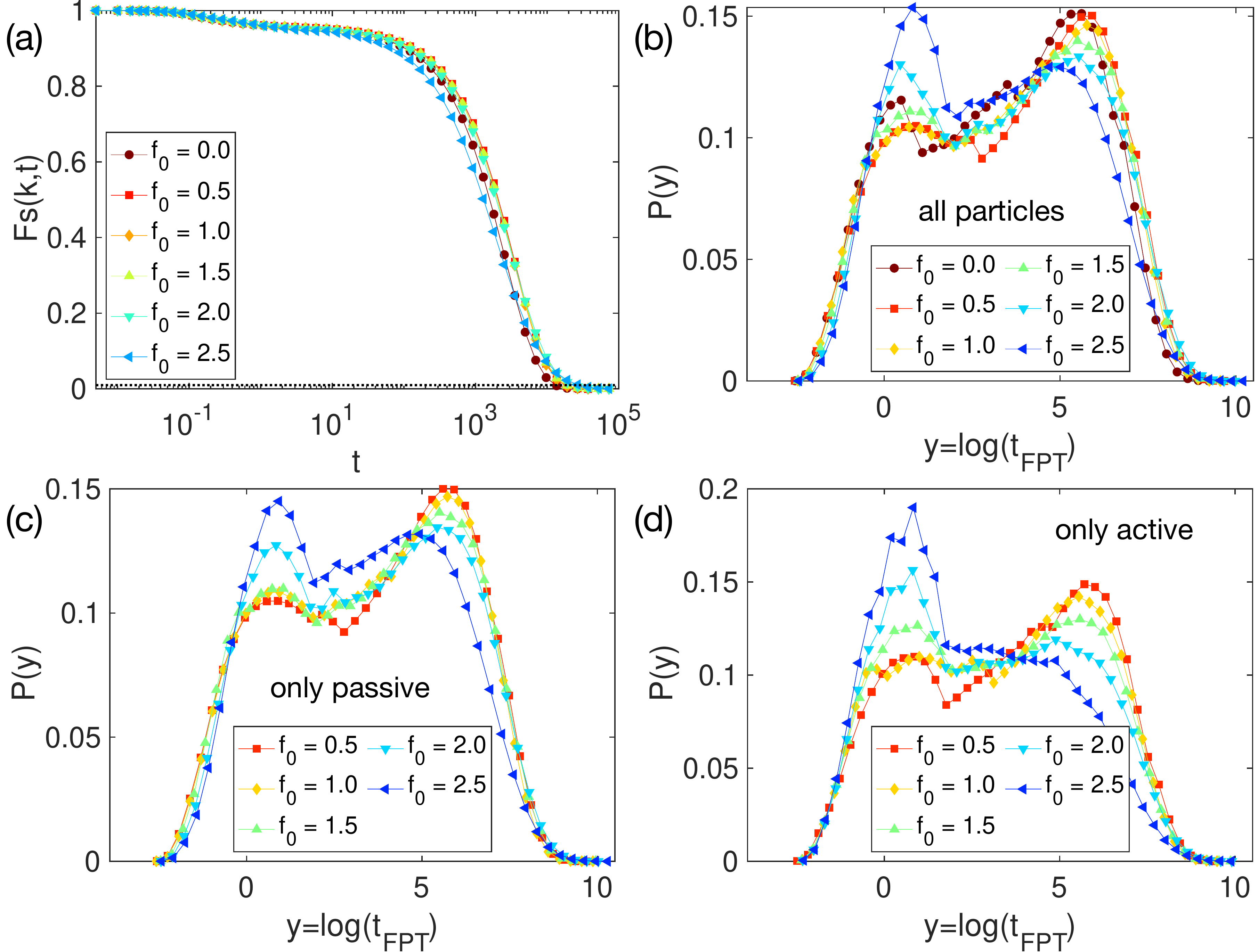}
\caption{(a)Relaxation profile of similarly relaxing active systems with different active forcing. The temperature of the system is so adjusted to match the relaxation times. (b) FPT distributions of the above mentioned systems; the increase in the fraction of hoppers with increasing activity is clear, thus the enhanced DH reported in \cite{PaulD2021,Mutneja2022}, and the increase in the power $\omega$.    }
\label{FPTSameTauAlpha}
\end{figure}
 Next, in Fig.\ref{FPTSameTauAlpha}(b), we show the FPT distribution of systems with different activities and bath temperature but having similar relaxation times as shown by the decay of $F_s(k,t)$ in Fig.\ref{FPTSameTauAlpha}(a). The increased amplitude of the first peak in Fig.\ref{FPTSameTauAlpha}(b) shows that the population of fast-moving hoppers increases with increasing activity, which is also supported by Fig.\ref{FPTSameTauAlpha}(c) and Fig.\ref{FPTSameTauAlpha}(d) where the same FPT distribution is computed by considering only the passive particles and only the active particles respectively. As argued earlier, these hoppers lead to the SE violation; thus, the increased population of hoppers with increasing activity will naturally mean stronger SE violation with a significant decrease in fractional SE power, $1-\omega$. This is also consistent with the observed increase in DH in a similar system by Paul et al. \cite{PaulD2021}. The FPT plots in supercooled regimes also support the fact of change in inherent dynamics of the constituents, from normal viscous to activated (dynamics of cages and hoppers) \cite{Faupel2003, Zhang2017, Mei2021}.

\begin{figure*}[htpb]
\includegraphics[width=0.95\textwidth]{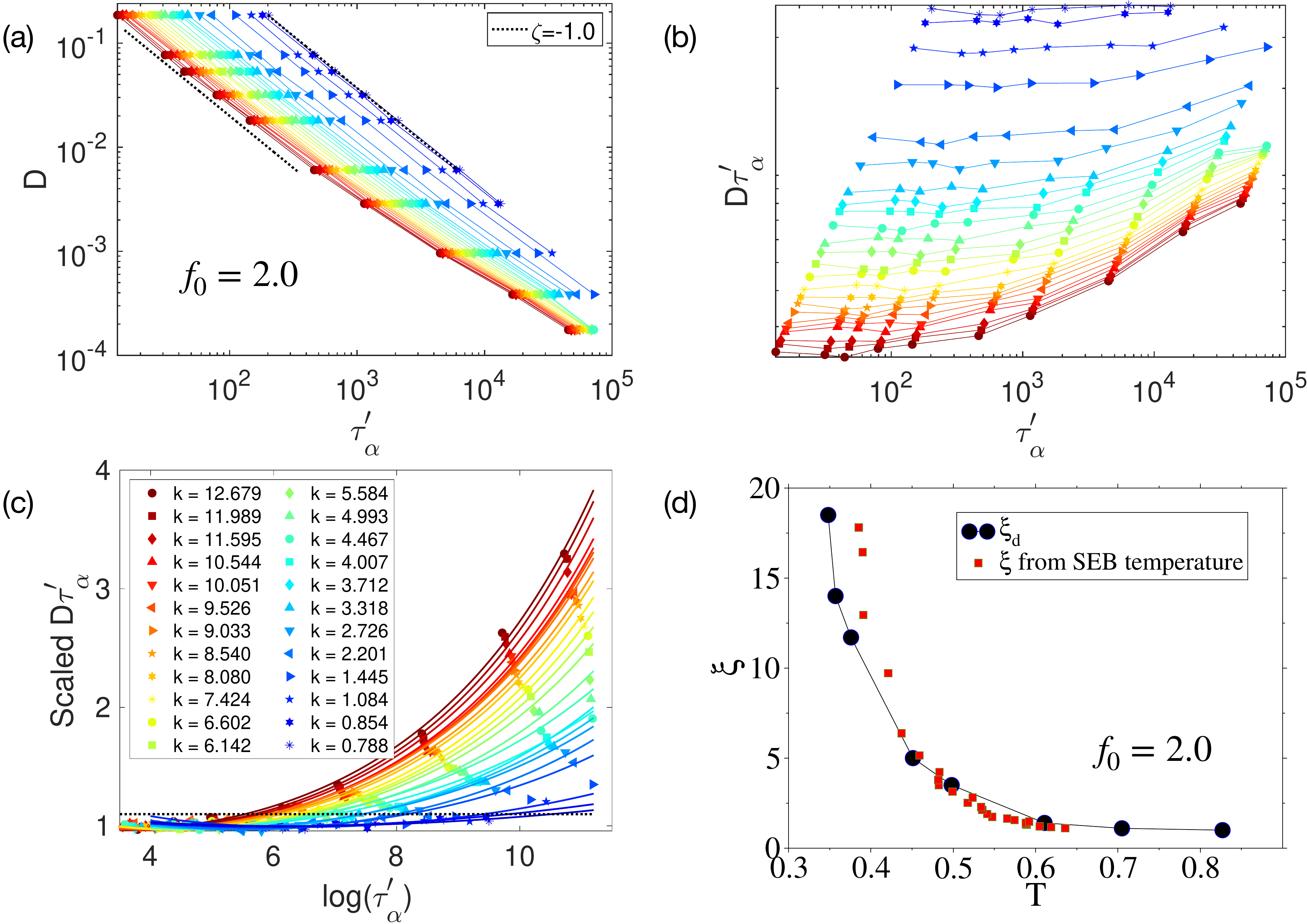}
\caption{(a) The cross plot of diffusion constant, $D$ against the relaxation time, $\tau_\alpha^\prime(k)$ for different wavevectors, $k$. The dotted line corresponds to $y=1/x$ and is obeyed for high temperatures and smaller $k$. (b) The SE parameter,  $D\tau_\alpha^\prime(k)$, is shown for different $k$. (c) The SE parameter scaled with the higher temperature values is shown. The breakdown happens at higher and higher temperatures for larger $k$. The lines in the plot are a fit to a function, $\ln{y}=x+c_0(x/x_0)^n+c_1(x/x_0)^{2n}$, where $x=log(\tau_\alpha^\prime)$, and the SEB temperature is calculated from the relaxation time for $y=1.1$ cut. (d) The length scale obtained from SE breakdown compares well with the $\xi_d$ obtained by conventional methods in Ref.\cite{PaulD2021}.}
\label{SEBF02.0}
\end{figure*}
We will now focus on examining the connection between the DH and the SE breakdown through the dynamic length scale, $\xi_d$. In equilibrium liquids, $\xi_d$ is responsible for regulating the finite size effects observed in the four-point dynamic susceptibility ($\chi_4^P(T, N)$), which measures the fluctuations in the two-point correlation function and serves as an indicator of the correlated relaxation process in a system \cite{KDSPNAS2009}. Similarly, it is found to be the same as the coarse-graining scale over which the vanHove function crosses over from non-Gaussian to Gaussian form \cite{Bhowmik2018}. $\xi_d$ also dictates the dependence of non-normal parameters for a probe particle in supercooled liquids on the size of the probe \cite{MutnejaR, MutnejaT}. The relaxation times obtained from the dynamics at length scales larger than the dynamic length scale of the system were also shown to be coupled to the diffusion constant in Ref.\cite{Parmar2017}. We establish a similar finding in nonequilibrium active systems too. Fig.\ref{SEBF02.0}(a) shows the log-log plot of diffusion constant ($D$) and relaxation time ($\tau^\prime_\alpha(k)$) for different wavevectors ($k$). Fig.\ref{SEBF02.0} is for a system with active forcing of magnitude $f_0=2.0$, while the results for all other active forcing are presented in Appendix. The relaxation times for wavevector $k$ can be thought of as the relaxation time probed at the coarse grain length scale of $ \sim \frac{2\pi}{k}$; thus, the larger $k$ values refer to the relaxation times probed at smaller length scales and vice versa. The relaxation times obtained for smaller $k$ values likely follow the SE relation, as seen from Fig. \ref{SEBF02.0} (b), where we plotted the SE parameter for different wavevectors. We scaled the SE parameter with its large temperature values \ref{SEBF02.0}(c) to observe the SE breakdown at different length scales systematically. The SE breakdown temperatures for different wavevectors are extracted by putting a threshold of $Scaled~D\tau^\prime_\alpha=1.1$. The SEB length scale compares well with the temperature dependence of the $\xi_d$ of the active system \cite{PaulD2021} as shown in Fig.\ref{FPTF02.0}(d). Filled circles refer to the data obtained in \cite{PaulD2021} via various methods of extracting the dynamical length, whereas the square symbols refer to the length scale obtained from the cross-over length scale above which SE is obeyed at that temperature and activity. This backs the hypothesis of the DH controlling the SEB in the active systems as well. Note that the dynamic length scale proliferates with supercooling in active glass-forming liquids and can grow by as much as $20 - 25$ particle diameter with increasing activity. A similar analysis for systems with different activities is provided in Appendix to highlight the generic nature of our results across changing activity. 

\section{Discussion}
To summarise, we have studied various aspects of SE relation in the active system with different activities. These systems are found to exhibit enhanced dynamic heterogeneity effects and large growing dynamic length scales. The active systems follow the SE relation in the limit of higher temperatures. However, the SE relation is violated at lower temperatures, and instead, the fractional SE relation is obeyed in complete accordance with passive supercooled liquids. The exponent of fractional SE relation however decreases systematically with increasing activity, and the exponent is found to decrease as much as $1-\omega=0.612$ when the relaxation time is computed at the probing wavevector of $k\sim7.25$ (near the peak of the structure factor) for the highest activity $f_0 = 2.5$. The exponent $\omega$ has significant wavevector dependence; it increases with increasing $k$ and reaches $1-\omega=0.573 $ for the highest activity and largest $k$ considered in this study. The first passage time (FPT) distributions of particles in these active supercooled liquids provided significant insight into enhanced SE breakdown in these systems. The FPT distributions developed a small time peak signifying the change in the inherent viscous dynamics to hopping-dominated dynamics with increasing activity. The fraction of hoppers were found to be larger in similarly relaxing but more active systems, which we conclude to be the main driving factor for the observed enhanced DH and stronger SE violation. The large time tail of FPT distributions was found to be scaled reasonably well by the relaxation time of the system, while the mean follows the diffusion coefficient. We showed that the breakdown temperature of the SE relation increase with increasing $k$, providing information on the intricate role played by growing dynamic length scales in the active systems. As the dynamic length scale in these systems can grow by nearly an order of magnitude compared to passive glass-forming liquids, the role of dynamic heterogeneity length scale in SE violation became even more prominent than their equilibrium counterparts. Finally, this work shows that the SE relation (an equilibrium concept) can be extended to inherently nonequilibrium systems to gain insights into the dynamics and to extract information on growing correlation lengths in the system. Although Stokes-Einstein breakdown in active liquids is found to be similar to the equilibrium supercooled liquids in the range of the studied parameters, it is not immediately clear whether these concepts can be extended for active systems in their extreme limit, for example, with very large persistence time $\tau_p$, at which the system can show intermittent dynamical behaviour \cite{Mandal2020}.  
\section{Acknowledgements}
We acknowledge funding by intramural funds at TIFR Hyderabad from the Department of Atomic Energy (DAE) under Project Identification No. RTI 4007. Core Research Grant CRG/2019/005373 from Science and Engineering Research Board (SERB) is acknowledged for generous funding. Most of the computations are done using the HPC clusters procured using CRG/2019/005373 grant and Swarna Jayanti Fellowship, grants DST/SJF/PSA01/2018-19, and SB/SFJ/2019-20/05.

\section{Appendix}
\begin{figure*}[!htpb]
\includegraphics[width=\textwidth]{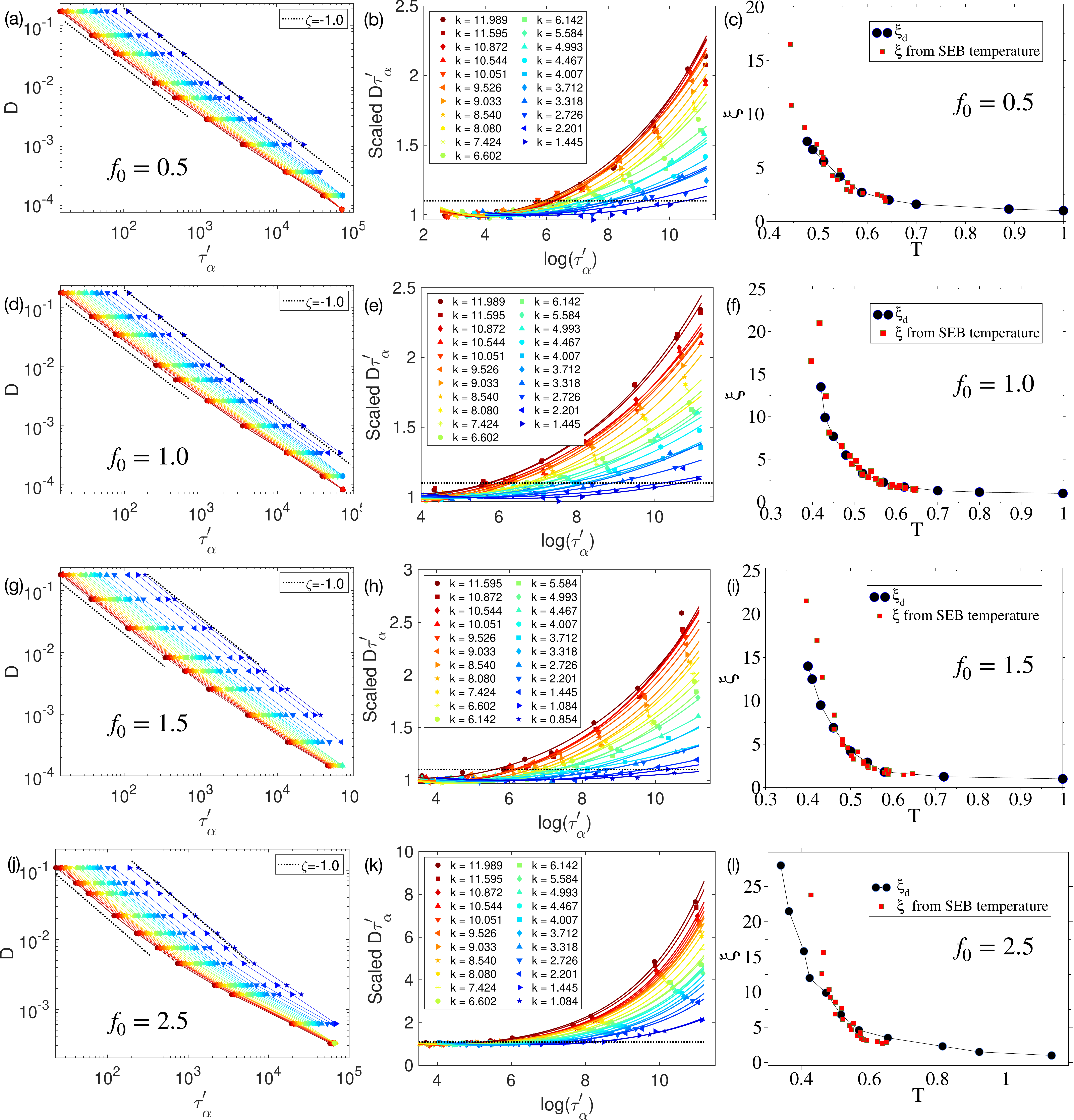}
\caption{The SE analysis to obtain the breakdown temperature for systems with different activity. See Results for details.}
\label{SEBAllActivities}
\end{figure*}

The SE breakdown analysis for different activities is presented in Fig.\ref{SEBAllActivities}. The length scale extracted compares well with the dynamic length scale of the system. 

\bibliography{activeWave}
\end{document}